\documentclass[preprint,3p,twocolumn]{elsarticle}
\usepackage{scrextend}
\usepackage{graphicx, epsfig,amssymb} 
\usepackage{amsmath, amsfonts}
\usepackage{bm}
\usepackage[breaklinks]{hyperref}
\usepackage{color}
\usepackage{enumerate}
\usepackage{amsmath}
\usepackage{cuted}
\usepackage{aas_macros}
\usepackage{capt-of}
\usepackage{adjustbox}
\usepackage{mathrsfs}

\definecolor{oucrimsonred}{rgb}{0.6, 0.0, 0.0}
\definecolor{persianblue}{rgb}{0.11, 0.22, 0.73}
\definecolor{forestgreen}{rgb}{0.13,0.35,0.13}
 \hypersetup{colorlinks, citecolor=oucrimsonred, linkcolor=persianblue, urlcolor=oucrimsonred}
 \hypersetup{colorlinks, citecolor=oucrimsonred, linkcolor=persianblue, urlcolor=oucrimsonred}

\usepackage{lmodern}

\newcommand{\gappeq}{{\rlap{{\raise}.5ex\text{\ensuremath{>}}}{{\lower}.5ex\text{\ensuremath{\sim}}}}}
\newcommand{\lappeq}{{\rlap{{\raise}.5ex\text{\ensuremath{<}}}{{\lower}.5ex\text{\ensuremath{\sim}}}}}

\newcommand{\I}{\tmtextrm{1{\kern}-.24em l}}

\def\be{\begin{equation}}
\def\ee{\end{equation}}
\newcommand{\bp}{\bar M_{Pl}}

\usepackage{lipsum}

\hypersetup{colorlinks, citecolor=bluscuro, linkcolor=black, urlcolor=bluscuro}
\definecolor{rossos}{cmyk}{0,1,1,0.55}
\definecolor{bluscuro}{rgb}{0.15, 0.2, .85}
\definecolor{bluchiaro}{cmyk}{1,.3,0.,0.1}
\usepackage{cancel}

\newcommand{\bea}{\begin{eqnarray}}
\newcommand{\eea}{\end{eqnarray}}
\newcommand{\bc}{\begin{center}}
\newcommand{\ec}{\end{center}}

\usepackage{adjustbox}
\usepackage{array}
\usepackage{booktabs}
\usepackage{multirow}

\newcolumntype{R}[2]{%
    >{\adjustbox{angle=#1,lap=\width-(#2)}\bgroup}%
    l%
    <{\egroup}%
}
\newcolumntype{R}[2]{%
    >{\adjustbox{angle=#1,lap=\width-(#2)}\bgroup}%
    l%
    <{\egroup}%
}

\def\be{\begin{equation}}
\def\ee{\end{equation}}
\def\ba{\begin{array} }
\def\bac{\begin{array} {c}}
\def\bacc{\begin{array} {cc}}
\def\baccc{\begin{array} {ccc}}
\def\bacccc{\begin{array} {cccc}}
\def\ea{\end{array}}
\def\bea{\begin{eqnarray}}
\def\eea{\end{eqnarray}}

\definecolor{red}{rgb}{1,0,0}

\def\psl{\hbox{\hbox{${p}$}}\kern-1.9mm{\hbox{${/}$}}}
\def\dsl{\hbox{\hbox{${\partial}$}}\kern-2.2mm{\hbox{${/}$}}}
\def\Dsl{\hbox{\hbox{${D}$}}\kern-2.6mm{\hbox{${/}$}}}

\def\Lag{\mathscr{L}}

\begin{document}
\title{ Radiative Axion Inflation}

\author{Andrea Caputo\fnref{fn1}}\ead{andrea.caputo@uv.es}\address{Instituto de F\'{i}sica Corpuscular, Universidad de Valencia and CSIC, Edificio Institutos Investigaci\'{o}n, Catedr\'{a}tico Jos\'{e} Beltr\'{a}n 2, 46980 Spain}
\begin{abstract} 
Planck data robustly exclude the simple $\lambda\phi^4$ scenario for inflation. 
This is also the case for models of "Axion Inflation" in which the inflaton field is the radial part of the Peccei-Quinn complex scalar field. In this letter we show that for the KSVZ model it is possible to match the data  taking into account radiative corrections to the tree level potential. After writing down the 1-loop Coleman-Weinberg potential, we show that a radiative plateau is easily generated thanks to the fact that the heavy quarks are charged under $SU(3)_c$ in order to solve the strong CP problem. We also give a numerical example for which the inflationary observables are computed and the heavy quarks are predicted to have a mass $m_Q \gtrsim\, 10^2\,\,TeV$.
\end{abstract}

\maketitle

\noindent{\bf{\em Introduction.}}
The nature of the inflaton field\cite{Linde:1981mu, Guth:1980zm} responsible to the initial acceleration of the universe remains unknown. In the recent years with the enormous amount of available cosmological data we have been facing the possibility to truly discriminate between different UV models. An intriguing possibility is that the field responsible of inflation is also able to solve some of the others issues we face in the standard model of particle physics. In this letter we study the possibility that the inflaton is the complex scalar field associated with the Peccei-Quinn symmetry; the presence of this field is justified to solve the strong CP problem and incidentally also generates a very promising dark matter candidate, the axion. The resulting model is a model of quartic inflation of the type $\lambda \phi^4$, which is well known to be disfavored by CMB data\cite{Akrami:2018odb} because of the too large tensor to scalar ratio predicted (in models of monomial inflation $\phi^n$ the larger is n the larger is the disagreement). A simple possibility to lower the tensor to scalar ratio r is found in model in which the inflaton couples to the Ricci scalar\cite{Bezrukov:2007ep, Libanov:1998wg, Fakir:1990eg, Futamase:1987ua, Masina:2018ejw,Okada:2010jf, Linde:1981mu, Kallosh:1995hi, Inagaki:2014wva}. This possibility, in the contest of "Axion Inflation", was already studied eg. in \cite{Ballesteros:2016xej, Fairbairn:2014zta}, where the authors showed that the non minimal coupling to gravity $\xi_{\phi}$ can flatten the potential at large values and reconcile predictions with data. \\However, another possibility to flatten the potential is given by radiative correction with fermionic loops, as showed in \cite{Ballesteros:2015noa, NeferSenoguz:2008nn, Enqvist:2013eua} (for other models for inflation and axions see e.g \cite{Odintsov:2019mlf,Grimm:2007hs, Bugaev:2013fya}). Here in particular we follow \cite{Ballesteros:2015noa} and show that the KSVZ model\cite{Kim:1979if, Shifman:1979if} has all the ingredients to rescue inflation. This is not the case instead for the DFSZ model\cite{Zhitnitsky:1980tq, Dine:1981rt}, in which there are no new fermionic degrees of freedom that couple to the inflaton. Of course a non-minimal coupling should be nevertheless included in a complete analysis, given that it is radiatively generated, even if it is set to zero at some scale. Nevertheless, the running of the non-minimal coupling depends on the scalar and Yukawa couplings of the theory (see $\beta_{\xi_A}$ in\ref{appendix}), which, as we will see, are small. The goal of this letter is to show that even with a negligible non-minimal coupling the model can fit the data. The considered model is a typical inflection-point inflation, which has already been considered in other contests (\cite{Okada:2016ssd, Okada:2017cvy, Choi:2016eif, Enqvist:2010vd}), here for the first time we considered it in a very well motivated scenario as the KSVZ model.\\
\noindent{\bf{\em The model.}}\\We consider the model with Lagrangian:
\be\mathscr{L} =  \Lag_{\rm gravity}+\Lag_{\rm SM}+  \Lag_{\rm a}  
,\label{full-lagrangian}
 \ee
The gauge group of the model is the SM one:  $$G_{\rm SM}= {\rm SU(3)_c\times SU(2)_{\it L}\times U(1)_{\it Y}}.$$ $\mathscr{L} _{\rm gravity}$ are the terms in the Lagrangian which include the pure gravitational part, which is $- \frac{\bp^2}{2} R -\Lambda$
 where $\bp\simeq 2.4\times 10^{18}\,$GeV is the reduced Planck mass and $\Lambda$ is the cosmological constant. 
  $\mathscr{L} _{\rm SM}$ is the SM Lagrangian.  
 $\mathscr{L} _{\rm a}$ represents the additional terms in the Lagrangian for the axion model. We consider the first invisible axion model, the so called KSVZ model. This model introduces new fields, namely:
\begin{itemize}
\item An extra Dirac fermion Q, colored but neutral under $\rm SU(2)_{\it L}\times U(1)_{\it Y}$, which in Weyl notation can be written as a pair of two-component fermions $q_1$ and $q_2$ in the following representation of $G_{\rm SM}$ 
\be q_1 \sim (3,1)_0, \qquad q_2 \sim (\bar{3}, 1)_0.\ee
\item An extra complex scalar $A$, charged under $\rm U(1)_{PQ}$ and neutral under $G_{\rm SM}$.
\end{itemize}
The Lagrangian of the KSVZ model is
\be \mathscr{L}_a =i\sum_{j=1}^2\overline{q}_j \Dsl \, q_j +|\partial_{\mu} A|^2  -(y\, q_2A q_1 +h.c.)-\Delta V(H,A)\nonumber \ee
and the classical potential reads
\be  V(H,A)=  \lambda_H\Big(|H|^2-\frac{v^2}{2}\Big)^2+\Delta V(H,A), \ee
where
\begin{multline}
 \Delta V(H,A) \equiv \lambda_A\Big(|A|^2-  \frac{f_a^2}{2}\Big)^2 + \\+ 2\lambda_{HA} \Big(|H|^2-\frac{v^2}{2}\Big)\Big( |A|^2-\frac{f_a^2}{2}\Big)
\end{multline}
The parameters $v$, $f_a$ and $y$ can be taken real and positive without loss of generality. The PQ symmetry acts on $q_1$, $q_2$ and $A$ as follows
\be q_1\rightarrow e^{i\alpha/2}q_1, \quad q_2\rightarrow e^{i\alpha/2}q_2, \quad A\rightarrow e^{-i\alpha}A, \ee
while the SM fields are instead neutral under U(1)$_{\rm PQ}$.
This model adds only three new real parameters: $\lambda_{HA}$, $\lambda_A$ and $y$. It is a very simple model with all the ingredients, as we will show in a moment, to generate a radiative plateau.\\
\noindent{\bf{\em Radiative Quartic Inflation.}} In the following we review the idea studied in \cite{Ballesteros:2015noa} which consists in the generation, through loop corrections, of a plateau in the inflaton potential which favors slow-roll inflation. A perfect plateau is characterized by (for simplicity we will indicated by A the radial part of the PQ field)
\begin{equation}
	\frac{dV}{dA}=0, \,\, \frac{d^2V}{dA^2}=0
	\label{der}
\end{equation}
In this study the role of the inflaton is played by the scalar field $A$, which is supposed to guide a typical scenario of chaotic inflation. 
For large values of the field the potential simplifies to 
\begin{equation}
	V(A) = \hat{\lambda}_A(A)A^4
	\label{effective}
\end{equation} 
where we notice we have already put field-dependent effective quartic coupling. The easiest way to understand this field dependence is to write down the Coleman-Weinberg form of the potential\cite{Coleman:1973jx, Barenboim:2013wra}
\begin{multline}
	V(A,\mu)=\lambda_A A^4+\\+\frac{1}{64\pi^2}\sum_i(-1)^{F_i}S_iM_i^4(A)\Big(ln\frac{M_i^2(A)}{\mu^2}-c_i\Big) +...
\end{multline}
where $\mu$ is the renormalization scale, $F_i=0(1)$ for a boson (fermion) field in the loop, $S_i$ counts the degrees of freedom of each particle (e.g 12 for a colored Dirac fermion) with a field-dependent mass $M_i(A,\mu)$, and $c_i=3/2$ for fermions and scalar bosons and $c_i=5/6$ for gauge bosons. The ellipsis stand for higher order loops. It is convenient to choose the renormalization scale $\mu$ as
\begin{equation}
	\mu \equiv \varepsilon A
\end{equation}
where $\varepsilon$ is a positive constant generically much smaller than one which comes from $M_i^2(A)\propto A^2$. From the form of the Coleman-Weinberg potential is easy understood that $\varepsilon$ parametrizes the smallness of the coupling of the inflaton to the other fields which enter the loops. For example, in our model the Higgs and the heavy quarks will enter the loops for the computation of the effective potential and therefore $\varepsilon$ can be chosen to be proportional to $\sqrt{\lambda_{HA}}\sim y$. Here we will consider only one effective mass scale; for the situation in which there are more than one the interested reader can look at \cite{Casas:1998cf}.Keeping only the terms containing the fourth power of A (which are the important ones during inflation), the logarithms are resummed into an effective quartic coupling as anticipated in \ref{effective}. Then, we can expand the effective quartic coupling $\hat{\lambda}_A$ around the location of the plateau $A_0$; the effective quartic coupling will be a combination of the nominal coupling of the theory which now run with the renormalization scale $\mu = \epsilon A$ (the theory is maintained fixed for a continuous change of the subtraction scale by the renormalization group).Expanding $\hat{\lambda}_A$ around $A_0$ therefore corresponds to make the nominal parameters running around the scale $\varepsilon A_0$. We can therefore define the beta-function (and its derivative) for the effective quartic coupling
\begin{equation}
	\beta_{\hat{\lambda}_A}\equiv \mu \frac{\partial\hat{\lambda}_A}{\partial\mu}, \,\,\beta'_{\hat{\lambda}_A}\equiv \mu\frac{\partial \beta_{\hat{\lambda}_A}}{\partial \mu}
\end{equation}
that are intended to be a combination of the beta functions of the Lagrangian parameters that define $\hat{\lambda}_A$. Consequently we can write
\begin{multline}
	\hat{\lambda}_A(A)=\hat{\lambda}_A(A_0) + \frac{1}{2}\beta_{\hat{\lambda}_A}|_{A_{0}}log\frac{A^2}{A_0^2} +\\+ \frac{1}{8}\beta'_{\hat{\lambda}_A}|_{A_{0}}\Big(log\frac{A^2}{A_0^2}\Big)^2 + \,\,....
\end{multline}
We stress again that by construction evaluating the effective quartic coupling $\hat{\lambda}_A$ at the scale $A_0$ corresponds to evaluate the original Lagrangian couplings at $\varepsilon A_0$.
The plateau conditions of Eq. \ref{der} can be written in terms of $\hat{\lambda}_A$ and its beta-function (and derivative of it)
\begin{equation}
	\beta_{\hat{\lambda}_A}|_{A_0} = -4\hat{\lambda}_A|_{A_0}, \,\, \beta'_{\hat{\lambda}_A}|_{A_0} = -4\beta_{\hat{\lambda}_A}|_{A_0}
	\label{plateau}
\end{equation}
and the effective potential assumes the form
\begin{equation}
	V(A)= \hat{\lambda}_A|_{A_0}\Big(1 -2log\frac{A^2}{A_0^2}+2\Big(log\frac{A^2}{A_0^2}\Big)^2\Big)A^4
\end{equation}
 \\Finally, since the potential has to be positive around the inflection point $A_0$, the effective quartic coupling has to be positive too. This will be the crucial thing to verify to check the viability of the model.\\\noindent{\bf{\em Plateau in the KSVZ model.}} Now we want to see if it is possible to obtain a plateau in this simple model. First of all we need the expression for the effective quartic coupling; the relation between the latter and the Lagrangian couplings is easily inferred from the Coleman-Weinberg potential
and for the KSVZ model is given by
\begin{multline}
	\hat{\lambda}_A= \lambda_A + 36\lambda_A^2k\Big(ln\frac{12\lambda_A}{\epsilon^2} -3/2 \Big) + \\4\lambda_{HA}^2k\Big(ln\frac{2\lambda_{HA}}{\epsilon^2} -3/2 \Big) - 3y^4k\Big( ln\frac{y^2}{\epsilon^2}  -3/2\Big)
	\label{effective}
\end{multline}
where we have defined the loop factor $k\equiv \frac{1}{16\pi^2}$

We will look for perturbative solutions of conditions \ref{plateau}, which can be solved to express two of the Lagrangian couplings in terms of the rest. We will solve the plateau condition for $\hat{\lambda}_A$ and $\lambda_{HA}$; to do it, it is necessary to know the effective quartic coupling and its beta-function. In the Appendix we report the relevant beta functions; the KSVZ model has been implemented and the RG equations calculated using SARAH \cite{Staub:2013tta}.\\
It is useful to make a formal expansion of the couplings of the Lagrangian and their beta functions, to take trace of the order. Labelling with $\delta_i$ a generic coupling, we can write
\begin{equation}
	\delta_i = \delta_i^{(0)} +k\delta_i^{(1)} +..
\end{equation}
\begin{equation}
	\beta_{\delta_i}\equiv \frac{\partial\delta_i}{\partial log\mu}= \beta_{\delta_i}^{(0)}+ k\beta_{\gamma_i}^{(1)}+..
\end{equation}
We then immediately notice that the equation
\begin{equation}
\beta'_{\hat{\lambda}_A}|_{A_0}=-4 \beta_{\hat{\lambda}_A}|_{A_0}
\end{equation}
 implies that the 1-loop contribution of the beta function of the effective quartic coupling has to be zero. This in turn implies ( because $\beta_{\hat{\lambda}_A}|_{A_0} = -4\hat{\lambda}_A|_{A_0}$) also that 
 \begin{equation}
 	 \hat{\lambda}_A^{(1)}=0
 \end{equation}
Moreover, we recall that at zero-th order $\hat{\lambda}_A^{(0)} = \lambda_A^{(0)}$ and that, because of the structure of \ref{effective}, it results 
\begin{equation}
	\beta_{\lambda_A}^{(1)}=\beta_{\hat{\lambda}_A}^{(1)} =0
\end{equation}
\\ We start imposing
\begin{equation}
	\beta_{\lambda_A}^{(1)}=\beta_{\hat{\lambda}_A}^{(1)}=0
\end{equation}
which in this model results in
 \begin{equation}
 	12\lambda_A^{(0)}y^2 -6y^4 +8(\lambda_{HA}^{(0)})^2 +20(\lambda_A^{(0)})^2 = 0
 \end{equation}
The zero-th order of the quartic coupling has to be zero at the plateau, because it is proportional to the beta function which is zero by definition at order $k^0$. Therefore
\begin{equation}
	(\lambda_{HA}^{(0)})^2 = \frac{3}{4}y^4
\end{equation} 
Then, we have to impose $\hat{\lambda}_A^{1}=0$ to find also $\lambda_A$ in terms of the rest of the parameters
\begin{multline}
	0=\hat{\lambda}_A^{1} = \lambda_A^{1} + 4(\lambda_{HA}^{(0)})^2 \Big(ln\frac{12\lambda_{HA}^{(0)}}{\epsilon^2} -3/2  \Big)+\\ -3y^4\Big(ln\frac{y^2}{\epsilon^2} -3/2\Big) 
	= \hat{\lambda}_A^{1l} + 3y^4 \Big(ln\frac{6\sqrt{3}y^2}{\epsilon^2} -3/2  \Big)+\\ -3y^4\Big(ln\frac{y^2}{\epsilon^2} -3/2\Big)
\end{multline}
We then finally get
\begin{equation}
	\lambda_A = -3ky^4 ln(6\sqrt{3}) 
\end{equation}
\\Then, the crucial point comes: we have to determine whether or not the effective quartic coupling at the plateau is positive. To do it we have to compute the first non-null contribution to the self coupling which results to be
\begin{equation}
	\hat{\lambda}_A^{(2)} = \frac{1}{16} \beta'^{(2)}_{\hat{\lambda}_A}
\end{equation}
and using the one loop beta functions and the conditions $\hat{\lambda}_A^{1}=0$ and $\beta_{\lambda_A}=\beta_{\hat{\lambda}_A}=0$ we get
\begin{multline}
	\hat{\lambda}_A=\frac{k}{16}\Big(-24y^3\beta_y^{(1)} +16\lambda_{HA}^{(0)}\beta_{\lambda_{HA}}^{1l}\Big)=\\= k\Big(-\frac{3}{4}y^3\beta_y^{(1)} +\frac{\sqrt{3}}{2}y^2\beta_{\lambda_{HA}}^{(1)} \Big)
\end{multline}
The involved beta functions at 1-loop are
\begin{multline}
	\beta_{\lambda_{HA}}=k\lambda_{HA}\Big(-\frac{9}{2}g_2^2 -\frac{9}{10}g_1^2 +8\lambda_{HA} +8\lambda_A \\+6(y^2+y_t^2+2\lambda_H)\Big)
\end{multline}
\begin{equation}
	\beta_{y}=4k\Big( y^3 -2g_3^2y \Big)
\end{equation}
and consequently the effective self coupling results
\begin{equation}
	\hat{\lambda}_A=k^2y^4\Big(6g_3^2 + y^2 +\sqrt{6}(y_t^2+2\lambda_H) -\sqrt{\frac{27}{8}} (g_2^2+ g_1^2 ) \Big)
\end{equation}
A crucial and peculiar point here is that the new fermions coupled to the inflaton are charged under $SU(3)_{color}$ in order to solve the CP strong problem. We can then replace the SM couplings with their values at a scale $\sim \varepsilon M_{Pl}$ and we see that the self coupling is always positive. For example
\begin{equation}
	\hat{\lambda}_A(A_0 \sim M_{Pl}) \sim k^2y^4(3.3 + y^2 )
\end{equation}
and we verified, using the RGEs of the standard model, that this is true for all the scales from $0.001M_{Pl}$ up to $30 M_{Pl}$, which is the interesting range to have a plateau in order to match CMB data. We recall that while the effective self coupling has to be evaluated at $A_0$, the nominal couplings of the theory have to be evaluated at $\varepsilon A_{0}$. As a reference value one can take $\varepsilon \sim 10^{-4}-10^{-1}$. We also notice that evaluating the SM couplings at the Planck scale using the RGEs of the SM is a good approximation as long as the new couplings are small. This is always the case for inflationary scenarios (see \cite{Ballesteros:2016xej}) where usually $\lambda_A \lessapprox 10^{-10}$.\\\noindent{\bf{\em A numerical example.}} We have just showed that the model admits a radiative plateau.
 We give here a numerical example and calculate the relevant inflationary quantities.
\begin{figure}[!htb!]
\centering
  \includegraphics[width=1.05\linewidth]{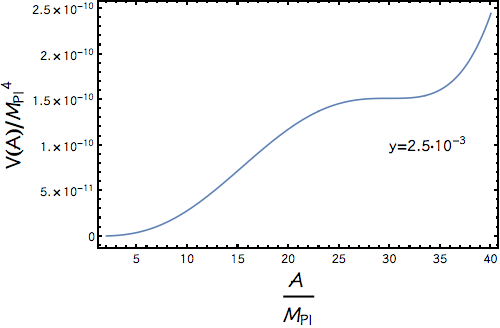}
  \vspace{-.1cm}
\caption{\label{fig:self}\em
Inflationary potential for $y=2.5\cdot10^{-3}$ and $ A_0=30\,M_{Pl}$}
\end{figure}
We fix the Yukawa coupling to be $y=2.5\cdot10^{-3}$ and the plateau location at $A_0=30\,M_{Pl}$; as mentioned above, all the other standard model couplings have been fixed at their values at $\varepsilon A_0$, with $\varepsilon \sim y =2.5\cdot 10^{-3}$. We then calculate the value of the field at which we match CMB observables and in particular the power spectrum
\begin{equation}
	\Delta_s^2 \sim \frac{1}{24\pi^2}\frac{V}{M_{Pl}^4\epsilon}\sim 2.142\cdot 10^{-9} \rightarrow A_i\sim 17M_{Pl}
\end{equation}
where 
\begin{equation}
	\epsilon=\frac{M_{Pl}^2}{2}\Big(\frac{V'}{V}\Big)^2
\end{equation}
We also compute the spectral index
\begin{equation}
	n_s \equiv \frac{dln\Delta_s^2}{dlnk}+1 \sim 2\eta -4\epsilon + 1
\end{equation}
and the tensor to scalar ratio
\begin{equation}
	r\equiv \frac{\Delta_t^2}{\Delta_s^2}=16\epsilon
\end{equation}
where $\Delta_t^2$ is the power spectrum of tensor fluctuations and $\eta$ is the second slow-roll parameter
\begin{equation}
	\eta= M_{Pl}^2\frac{V''}{V}
\end{equation}
At this value of the field the tensor to scalar ratio results
\begin{equation}
	r=0.079
\end{equation}
while the spectral index is
\begin{equation}
	n_s=0.963
\end{equation}
both consistent with the latest Planck release combining temperature, low-polarization, and lensing\cite{Akrami:2018odb}. 
\begin{figure}[!htb!]
\centering
\includegraphics[width=1.0\linewidth]{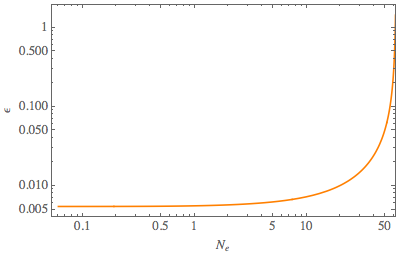}
  \vspace{-.1cm}
\caption{\label{fig:self}\em
Evolution of the first slow-roll parameter during inflation}
\label{epsilon}
\end{figure}

Then, we have to verify inflation lasts long enough in order to solve the flatness and horizon problems. Therefore we solve the equation of motion of the field as a function of the number of e-folds. In this way we do not rely on the slow-roll approximation and can precisely determine the number of e-folds $N_e$ at the end of inflation\cite{Ballesteros:2014yva}. In Fig.\ref{epsilon} we show the evolution of the slow-roll parameter $\epsilon$ as a function of the number of e-folds. Inflation ends when $\epsilon \sim 1$, which in this case happens for
\begin{equation}
 N_e \sim 64
\end{equation}
 a standard and satisfactory value. 
 
 \begin{figure}[!htb!]
\centering
\includegraphics[width=1.0\linewidth]{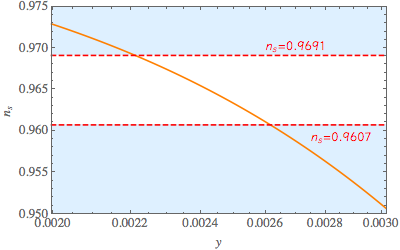}
  \vspace{-.1cm}
\caption{\label{fig:self}\em
Spectral index as a function of the Yukawa coupling. The two dashed red lines correspond to the experimental upper and lower bounds at $1\sigma$, respectively $n_s=0.9691$ and $n_s=0.9607$. The light blue regions are excluded by Planck\cite{Akrami:2018odb}.}
\label{nsplot}
\end{figure}

 \begin{figure}[!htb!]
\centering
\includegraphics[width=1.0\linewidth]{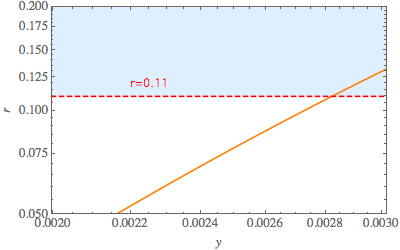}
  \vspace{-.1cm}
\caption{\label{fig:self}\em
Tensor to scalar ration as a function of the Yukawa coupling. The dashed red line corresponds to $r=0.11$. The light blue region is excluded at $95\% C.L$ by combining Planck temperature, low-polarization, and lensing\cite{Akrami:2018odb}.}
\label{rplot}
\end{figure}
We also explore the tuning of the model and show the tensor to scalar ratio and the spectral index as a function of the Yukawa coupling y, varying it around the referred value $y=2.5 \cdot 10^{-3}$(see Fig.\ref{nsplot}-\ref{rplot}) and keeping fixed the plateau position.\\ We notice that for 
 \begin{equation}
  2.2\cdot 10^{-3} \lesssim  y \lesssim 2.8 \cdot 10^{-3}
  \end{equation}
the radiative plateau fits the CMB data; in particular slightly changing the Yukawa coupling, the tensor to scalar ratio becomes consistent also with the stronger bound from the joint cross-correlation between Planck and BICEP 
\begin{equation}
	r < 0.064 \,(0.075)\,\,\, 95\% C.L
\end{equation}
using Plik (CAMspec) as high-l TT, TE, EE likelihood. We also verified that for all these values the number of e-folds spans a reasonable range $ 42 \lesssim N_e \lesssim 65$.
 Once the Yukawa coupling is fixed the mass of the heavy quark at low scale is given by
\begin{equation}
	m_{Q}=y\cdot f_a
\end{equation}
where $f_a$ is the breaking scale of the Peccei-Quinn symmetry and the Yukawa has to be evaluated at the desired scale. The lower bound for $f_a$ coming from Supernova cooling\cite{Raffelt:1999tx}
\begin{equation}
	f_a \gtrsim 4\cdot 10^8 GeV
\end{equation}
therefore implies
\begin{equation}
	m_{Q} \gtrsim 10^{5}GeV = 10^2\,\, TeV
\end{equation}\\
This is an important point because heavy relics could overclose if $M_{heavy}\gtrapprox 240 TeV$\cite{Griest:1989wd}. In the model considered here the Yukawa couplings can be even smaller than the considered benchmark value and therefore even smaller heavy quarks masses are obtained. This is indeed possible due to the structure of the beta functions. Instead this is not the case for other simple U(1) extensions where the Yukawa's required to generate a plateau are much larger (see e.g \cite{Ballesteros:2015noa}) and would produce dangerous heavy relics. 

\noindent{\bf{\em Baryogenesis and reheating.}}
Any inflationary scenario should be connected to Standard Big Bang Cosmology; this happens through the reheating phase, when the inflaton oscillates around the minimum of its potential and decay to populate the universe. In our scenario the reheating would proceed in a way similar to the one considered in \cite{Ballesteros:2016xej}, although with important differences. In fact, in \cite{Ballesteros:2016xej} the authors were trying to address other problems of the SM such as Higgs Instability, Dark Matter and Baryogenesis; this notably restricted the allowed parameter space.A more general study is underway and the results of numerical simulations will be presented elsewhere\footnote{G.Ballesteros, A.Caputo and C.Tamarit in preparation}.Preliminary results show that is indeed possible to efficiently reheat the universe in this model. Finally, the generation of matter-antimatter asymmetry is another compelling problem which also \cite{Ballesteros:2016xej} was trying to address. However inflationary physics and baryogenesis can be quite uncorrelated; for example, this would be the case in the presence of sterile neutrinos at the GeV scale. For Majorana neutrinos in the 1-100 GeV range, it has been shown by Akhmedov, Rubakov
and Smirnov (ARS) \cite{Akhmedov:1998qx} and refined by Asaka and Shaposhnikov (AS) \cite{Asaka:2005pn} that a peculiar mechanism of leptogenesis is at work. In this case the asymmetries are produced at freeze-in of the sterile states via their CP-violating oscillations. This scenario has been extensively studied \cite{Asaka:2005pn,Shaposhnikov:2008pf,Canetti:2012zc,Canetti:2012kh,Asaka:2011wq,Shuve:2014zua,Abada:2015rta,Hernandez:2015wna,Hernandez:2016kel,Drewes:2016gmt,Drewes:2016jae,Hambye:2016sby,Ghiglieri:2017gjz,Asaka:2017rdj,Hambye:2017elz,Abada:2017ieq,Ghiglieri:2017csp} and leads to the interesting possibility that the extra heavy neutrinos could be produced and searched for in beam dump experiments and colliders (see \cite{Ferrari:2000sp,Graesser:2007pc,delAguila:2008cj,BhupalDev:2012zg,Helo:2013esa,Blondel:2014bra,Abada:2014cca,Cui:2014twa,Antusch:2015mia,Gago:2015vma,Antusch:2016vyf,Caputo:2016ojx,Caputo:2017pit} for an incomplete list of works). A GeV scale for the sterile neutrinos implies very small Yukawa couplings $Y_{sterile} \lessapprox 10^{-6}-10^{-7}$, which therefore won't affect our results for the generation of the radiative plateau. 
\\
\noindent{\bf{\em Discussion.}} In this letter we analyzed an inflationary scenario in which the inflaton field is the scalar field which breaks the Peccei-Quinn symmetry in the KSVZ model.\,In particular we showed that radiative corrections are enough in order to match CMB observables and solve the flatness and horizon problems, with no need to resort to a non-minimal coupling to gravity. After a general discussion about the possibility to generate a plateau in the KSVZ model, we gave a numerical example which satisfies all cosmological constraints. Incidentally, we also show (see the appendix) all the beta functions at 2-loops of the minimal KSVZ model, even if, for the desired order of precision, the beta functions and the effective potential at 1-loop were sufficient in the numerical analysis. The necessary ingredients to have a radiative plateau are the scalar and fermions coupled to the inflaton. The peculiar ingredient of the KSVZ is the fact that the new fermions are charged under $SU(3)_c$ in order to solve the strong CP problem. This  avoids the need of large Yukawa couplings and makes the generation of the plateau natural. Because of this, as already stated, another well know axion model, namely the DFSZ model \cite{Zhitnitsky:1980tq, Dine:1981rt}, cannot lead to the generation of a radiative plateau.\,Nevertheless we want to mention that there are motivated extensions where the inflaton is coupled to other fermions. For example, in \cite{Langacker:1986rj, Caputo:2018zky, Clarke:2015bea} the authors consider the DFSZ model extended with sterile neutrinos coupled to the Peccei-Quinn field in order to explain the smallness of active neutrinos masses and baryogenesis.\,We expect a radiative plateau can be generated and inflation obtained also in those extensions. 
\\\noindent{\bf{\em Acknowledgements.}}We are warmly grateful to Guillermo Ballesteros, Pilar Hernandez, Mario Reig, Laura Sberna, Carlos Tamarit, Marco Taoso and Alfredo Urbano for reading and commenting the manuscript. We also thank Isabella Masina for discussions.\\
This work was partially supported by grants FPA2014-57816-P, PROMETEOII/2014/050 and SEV-2014-0398,  as well as by  the EU projects H2020-MSCA-RISE-2015 and H2020-MSCA-ITN-2015//674896-ELUSIVES. 
\\\\
A special thank to Mariagiovanna Malara for her support. To her I dedicate not only this work (it may not be that useful) but my all.

\begin{strip}
\section*{Appendix}
\label{appendix}
We show here the beta-functions of the model; for the fermions of the standard model we display only the Yukawa of the top $y_t$.\\
\begin{equation}
	\beta_{g_1}=\frac{1}{16\pi^2}\frac{41g_1^3}{10} +\frac{1}{256\pi^4}\frac{g_1^3(199g_1^2+135g_2^2+440g_3^2-85y_t^2)}{50}
\end{equation}
\begin{equation}
	\beta_{g_2}=-\frac{1}{16\pi^2}\frac{19g_2^3}{6} + \frac{1}{256\pi^4}\frac{g_2^3(27g_1^2+175g_2^2+360g_3^2-45y_t^2)}{30}
\end{equation} 
 \begin{equation}
 	\beta_{g_3}=-\frac{1}{16\pi^2}5g_3^3 +\frac{1}{256\pi^4}\frac{g_3^3(11g_1^2+45g_2^2+120g_3^2-10y^2 - 20y_t^2)}{10}
 \end{equation}
 \begin{align}
 	\beta_{y_t}=\frac{1}{16\pi^2}\Big(\frac{3 y_t^3}{2}+y_t(-\frac{17 g_1^2}{20}-\frac{9g_2^2}{4}-8g_3^2+3y_t^2)\Big)+\frac{1}{256\pi^4}\Big(y_t\Big(\frac{17}{8}g_1^2y_t^2+\frac{45}{8}g_2^2y_t^2 +20g_3^2y_t^2 -\frac{27}{4}y_t^4 -\frac{9}{20}g_1^2g_2^2 +\frac{19}{15}g_1^2g_3^2 +\\+\frac{1187}{600}g_1^4+9g_2^2g_3^2-\frac{23g_2^4}{4}-\frac{284g_3^4}{3}+2\lambda_{HA}^2+6\lambda_H^2\Big)+\frac{1}{80}\Big(120y_t^5 +y_t^3(-540y_t^2+223g_1^2+675g_2^2+1280 g_3^2-960\lambda_H)\Big)\Big)
\end{align}
\begin{equation}
\beta_y=\frac{1}{16\pi^2}4(y^3-2g_3^2y)+\frac{1}{256\pi^4}\Big(y\Big(20g_3^2y^2 -\frac{9}{2}y^4-\frac{284}{3}g_3^4+4\lambda_{HA}^2+4\lambda_A^2     \Big)  +y^3\Big(-\frac{9}{2}y^2 +\frac{32g_3^2}{3} -8\lambda_A\Big) +\frac{7}{4}y^5  \Big) 	
\end{equation}

\begin{multline}
	\beta_{\lambda_H}=\frac{1}{16\pi^2}\Big(12y_t^2 -6y_t^4+\frac{9}{20}g_1^2g_2^2 -\frac{9}{5}g_1^2\lambda_H +\frac{27}{200}g_1^4 -9g_2^2\lambda_H +\frac{9g_2^4}{8} +4\lambda_{HA}^2 +24\lambda_H  \Big) +\\ +\frac{1}{256\pi^2}\Big(\frac{63}{10}g_1^2g_2^2y_t^2 +\frac{17}{2}g_1^2\lambda_Hy_y^2 -\frac{171}{100}g_1^4y_t^2-\frac{8}{5}g_1^2y_t^4 +\frac{45}{2}g_2^2\lambda_Hy_t^2 -\frac{9}{4}g_2^2y_t^2 +80g_3^2\lambda_Hy_t^2 -32g_3^2y_t^4 -24\lambda_{HA}^2y^2 -3\lambda_Hy_t^4 -144\lambda_H y_t^2 +\\+30y_t^6+\frac{117}{20}g_1^2g_2^2\lambda_H -\frac{1677}{400}g_1^4g_2^2 -\frac{289}{80}g_1^2g_2^4 +\frac{1887}{200}g_1^4\lambda_H +\frac{108}{5}g_1^2\lambda_H^2 -\frac{3411g_1^6}{2000} +108g_2^2\lambda_H^2 -\frac{73}{8}g_2^4\lambda_H +\frac{305g_2^6}{16} -40\lambda_{HA}+\\ -32\lambda_{HA}^3 -312\lambda_H^3    \Big)
\end{multline}
\begin{multline}
	\beta_{\lambda_{HA}}=\frac{1}{16\pi^2}\frac{\lambda_{HA}\Big(60y^2 +60y_t^2 -9g_1^2 -45g_2^2 +80\lambda_{HA} +80\lambda_A +120\lambda_H  \Big)}{10}+\\ +\frac{1}{256\pi^4}\Big(\frac{17}{4}g_1^2\lambda_{HA}y_t^2 +\frac{45}{4}g_2^2\lambda_{HA}y_t^2 +40g_3^2\lambda_{HA}y^2 +40g_3^2\lambda_{HA}y_t^2 -48\lambda_{HA}\lambda_Ay^2 -9\lambda_{HA}y^4 -24\lambda_{HA}^2y^2 -72\lambda_{HA}\lambda_Hy_t^2 -\frac{27}{2}\lambda_{HA}y_t^4 +\\+24\lambda_{HA}^2y_t^2 +\frac{9}{8}g_1^2g_2^2\lambda_{HA}+\frac{72}{5}g_1^2\lambda_{HA}\lambda_H +\frac{1671}{400}g_1^4\lambda_{HA}+\frac{6}{5}g_1^2\lambda_{HA}^2 +72g_2^2\lambda_{HA}\lambda_H +6g_2^2\lambda_{HA}^2 -\frac{145}{16}g_2^4\lambda_{HA} -40\lambda_{HA}\lambda_A^2 -96\lambda_{HA}^2\lambda_A+ \\-60\lambda_{HA}\lambda_H^2 -144\lambda_{HA}\lambda_H -44\lambda_{HA}^3     \Big)
\end{multline}
\begin{equation}
	\beta_{\xi_A}=\frac{1}{16\pi^2}\Big((1+6\xi)\Big(\frac{y^2}{2}+\frac{2}{3}\lambda_A\Big)-\frac{\lambda_{HA}}{3}\Big)
\end{equation}
\begin{multline}
	\beta_{\lambda_A}=\frac{1}{16\pi^2}\Big(12\lambda_Ay^2 -6y^4 +8\lambda_{HA}^2 +20\lambda_A^2   \Big) +\frac{1}{256\pi^4}\Big(80g_3^2\lambda_Ay^2 -32g_3^2y^4 -48\lambda_{HA}^2y_t^2 +6\lambda_Ay^4 -120\lambda_A^2y^2 +24y^6+\frac{48}{5}g_1^2\lambda_{HA}^2+\\+48g_2^2\lambda_{HA}^2-80\lambda_{HA}^2\lambda_A-64\lambda_{HA}^3 -240\lambda_A^3 \Big)
\end{multline}

\end{strip}
 
\bibliography{axion}

\begin{thebibliography}{68}
\expandafter\ifx\csname natexlab\endcsname\relax\def\natexlab#1{#1}\fi
\expandafter\ifx\csname bibnamefont\endcsname\relax
  \def\bibnamefont#1{#1}\fi
\expandafter\ifx\csname bibfnamefont\endcsname\relax
  \def\bibfnamefont#1{#1}\fi
\expandafter\ifx\csname citenamefont\endcsname\relax
  \def\citenamefont#1{#1}\fi
\expandafter\ifx\csname url\endcsname\relax
  \def\url#1{\texttt{#1}}\fi
\expandafter\ifx\csname urlprefix\endcsname\relax\def\urlprefix{URL }\fi
\providecommand{\bibinfo}[2]{#2}
\providecommand{\eprint}[2][]{\url{#2}}

\bibitem[{\citenamefont{Linde}(1982)}]{Linde:1981mu}
\bibinfo{author}{\bibfnamefont{A.~D.} \bibnamefont{Linde}},
  \bibinfo{journal}{Phys. Lett.} \textbf{\bibinfo{volume}{108B}},
  \bibinfo{pages}{389} (\bibinfo{year}{1982}), \bibinfo{note}{[Adv. Ser.
  Astrophys. Cosmol.3,149(1987)]}.

\bibitem[{\citenamefont{Guth}(1981)}]{Guth:1980zm}
\bibinfo{author}{\bibfnamefont{A.~H.} \bibnamefont{Guth}},
  \bibinfo{journal}{Phys. Rev.} \textbf{\bibinfo{volume}{D23}},
  \bibinfo{pages}{347} (\bibinfo{year}{1981}), \bibinfo{note}{[Adv. Ser.
  Astrophys. Cosmol.3,139(1987)]}.

\bibitem[{\citenamefont{Akrami et~al.}(2018)}]{Akrami:2018odb}
\bibinfo{author}{\bibfnamefont{Y.}~\bibnamefont{Akrami}} \bibnamefont{et~al.}
  (\bibinfo{collaboration}{Planck}) (\bibinfo{year}{2018}),
  \eprint{1807.06211}.

\bibitem[{\citenamefont{Bezrukov and Shaposhnikov}(2008)}]{Bezrukov:2007ep}
\bibinfo{author}{\bibfnamefont{F.~L.} \bibnamefont{Bezrukov}} \bibnamefont{and}
  \bibinfo{author}{\bibfnamefont{M.}~\bibnamefont{Shaposhnikov}},
  \bibinfo{journal}{Phys. Lett.} \textbf{\bibinfo{volume}{B659}},
  \bibinfo{pages}{703} (\bibinfo{year}{2008}), \eprint{0710.3755}.

\bibitem[{\citenamefont{Libanov et~al.}(1998)\citenamefont{Libanov, Rubakov,
  and Tinyakov}}]{Libanov:1998wg}
\bibinfo{author}{\bibfnamefont{M.~V.} \bibnamefont{Libanov}},
  \bibinfo{author}{\bibfnamefont{V.~A.} \bibnamefont{Rubakov}},
  \bibnamefont{and} \bibinfo{author}{\bibfnamefont{P.~G.}
  \bibnamefont{Tinyakov}}, \bibinfo{journal}{Phys. Lett.}
  \textbf{\bibinfo{volume}{B442}}, \bibinfo{pages}{63} (\bibinfo{year}{1998}),
  \eprint{hep-ph/9807553}.

\bibitem[{\citenamefont{Fakir and Unruh}(1990)}]{Fakir:1990eg}
\bibinfo{author}{\bibfnamefont{R.}~\bibnamefont{Fakir}} \bibnamefont{and}
  \bibinfo{author}{\bibfnamefont{W.~G.} \bibnamefont{Unruh}},
  \bibinfo{journal}{Phys. Rev.} \textbf{\bibinfo{volume}{D41}},
  \bibinfo{pages}{1783} (\bibinfo{year}{1990}).

\bibitem[{\citenamefont{Futamase and Maeda}(1989)}]{Futamase:1987ua}
\bibinfo{author}{\bibfnamefont{T.}~\bibnamefont{Futamase}} \bibnamefont{and}
  \bibinfo{author}{\bibfnamefont{K.-i.} \bibnamefont{Maeda}},
  \bibinfo{journal}{Phys. Rev.} \textbf{\bibinfo{volume}{D39}},
  \bibinfo{pages}{399} (\bibinfo{year}{1989}).

\bibitem[{\citenamefont{Masina}(2018)}]{Masina:2018ejw}
\bibinfo{author}{\bibfnamefont{I.}~\bibnamefont{Masina}},
  \bibinfo{journal}{Phys. Rev.} \textbf{\bibinfo{volume}{D98}},
  \bibinfo{pages}{043536} (\bibinfo{year}{2018}), \eprint{1805.02160}.

\bibitem[{\citenamefont{Okada et~al.}(2010)\citenamefont{Okada, Rehman, and
  Shafi}}]{Okada:2010jf}
\bibinfo{author}{\bibfnamefont{N.}~\bibnamefont{Okada}},
  \bibinfo{author}{\bibfnamefont{M.~U.} \bibnamefont{Rehman}},
  \bibnamefont{and} \bibinfo{author}{\bibfnamefont{Q.}~\bibnamefont{Shafi}},
  \bibinfo{journal}{Phys. Rev.} \textbf{\bibinfo{volume}{D82}},
  \bibinfo{pages}{043502} (\bibinfo{year}{2010}), \eprint{1005.5161}.

\bibitem[{\citenamefont{Kallosh et~al.}(1995)\citenamefont{Kallosh, Linde,
  Linde, and Susskind}}]{Kallosh:1995hi}
\bibinfo{author}{\bibfnamefont{R.}~\bibnamefont{Kallosh}},
  \bibinfo{author}{\bibfnamefont{A.~D.} \bibnamefont{Linde}},
  \bibinfo{author}{\bibfnamefont{D.~A.} \bibnamefont{Linde}}, \bibnamefont{and}
  \bibinfo{author}{\bibfnamefont{L.}~\bibnamefont{Susskind}},
  \bibinfo{journal}{Phys. Rev.} \textbf{\bibinfo{volume}{D52}},
  \bibinfo{pages}{912} (\bibinfo{year}{1995}), \eprint{hep-th/9502069}.

\bibitem[{\citenamefont{Inagaki et~al.}(2014)\citenamefont{Inagaki, Nakanishi,
  and Odintsov}}]{Inagaki:2014wva}
\bibinfo{author}{\bibfnamefont{T.}~\bibnamefont{Inagaki}},
  \bibinfo{author}{\bibfnamefont{R.}~\bibnamefont{Nakanishi}},
  \bibnamefont{and} \bibinfo{author}{\bibfnamefont{S.~D.}
  \bibnamefont{Odintsov}}, \bibinfo{journal}{Astrophys. Space Sci.}
  \textbf{\bibinfo{volume}{354}}, \bibinfo{pages}{2108} (\bibinfo{year}{2014}),
  \eprint{1408.1270}.

\bibitem[{\citenamefont{Ballesteros et~al.}(2017)\citenamefont{Ballesteros,
  Redondo, Ringwald, and Tamarit}}]{Ballesteros:2016xej}
\bibinfo{author}{\bibfnamefont{G.}~\bibnamefont{Ballesteros}},
  \bibinfo{author}{\bibfnamefont{J.}~\bibnamefont{Redondo}},
  \bibinfo{author}{\bibfnamefont{A.}~\bibnamefont{Ringwald}}, \bibnamefont{and}
  \bibinfo{author}{\bibfnamefont{C.}~\bibnamefont{Tamarit}},
  \bibinfo{journal}{JCAP} \textbf{\bibinfo{volume}{1708}}, \bibinfo{pages}{001}
  (\bibinfo{year}{2017}), \eprint{1610.01639}.

\bibitem[{\citenamefont{Fairbairn et~al.}(2015)\citenamefont{Fairbairn, Hogan,
  and Marsh}}]{Fairbairn:2014zta}
\bibinfo{author}{\bibfnamefont{M.}~\bibnamefont{Fairbairn}},
  \bibinfo{author}{\bibfnamefont{R.}~\bibnamefont{Hogan}}, \bibnamefont{and}
  \bibinfo{author}{\bibfnamefont{D.~J.~E.} \bibnamefont{Marsh}},
  \bibinfo{journal}{Phys. Rev.} \textbf{\bibinfo{volume}{D91}},
  \bibinfo{pages}{023509} (\bibinfo{year}{2015}), \eprint{1410.1752}.

\bibitem[{\citenamefont{Ballesteros and Tamarit}(2016)}]{Ballesteros:2015noa}
\bibinfo{author}{\bibfnamefont{G.}~\bibnamefont{Ballesteros}} \bibnamefont{and}
  \bibinfo{author}{\bibfnamefont{C.}~\bibnamefont{Tamarit}},
  \bibinfo{journal}{JHEP} \textbf{\bibinfo{volume}{02}}, \bibinfo{pages}{153}
  (\bibinfo{year}{2016}), \eprint{1510.05669}.

\bibitem[{\citenamefont{Senoguz and Shafi}(2008)}]{NeferSenoguz:2008nn}
\bibinfo{author}{\bibfnamefont{V.~N.} \bibnamefont{Senoguz}} \bibnamefont{and}
  \bibinfo{author}{\bibfnamefont{Q.}~\bibnamefont{Shafi}},
  \bibinfo{journal}{Phys. Lett.} \textbf{\bibinfo{volume}{B668}},
  \bibinfo{pages}{6} (\bibinfo{year}{2008}), \eprint{0806.2798}.

\bibitem[{\citenamefont{Enqvist and Karciauskas}(2014)}]{Enqvist:2013eua}
\bibinfo{author}{\bibfnamefont{K.}~\bibnamefont{Enqvist}} \bibnamefont{and}
  \bibinfo{author}{\bibfnamefont{M.}~\bibnamefont{Karciauskas}},
  \bibinfo{journal}{JCAP} \textbf{\bibinfo{volume}{1402}}, \bibinfo{pages}{034}
  (\bibinfo{year}{2014}), \eprint{1312.5944}.

\bibitem[{\citenamefont{Odintsov and Oikonomou}(2019)}]{Odintsov:2019mlf}
\bibinfo{author}{\bibfnamefont{S.~D.} \bibnamefont{Odintsov}} \bibnamefont{and}
  \bibinfo{author}{\bibfnamefont{V.~K.} \bibnamefont{Oikonomou}},
  \bibinfo{journal}{Phys. Rev.} \textbf{\bibinfo{volume}{D99}},
  \bibinfo{pages}{064049} (\bibinfo{year}{2019}), \eprint{1901.05363}.

\bibitem[{\citenamefont{Grimm}(2008)}]{Grimm:2007hs}
\bibinfo{author}{\bibfnamefont{T.~W.} \bibnamefont{Grimm}},
  \bibinfo{journal}{Phys. Rev.} \textbf{\bibinfo{volume}{D77}},
  \bibinfo{pages}{126007} (\bibinfo{year}{2008}), \eprint{0710.3883}.

\bibitem[{\citenamefont{Bugaev and Klimai}(2014)}]{Bugaev:2013fya}
\bibinfo{author}{\bibfnamefont{E.}~\bibnamefont{Bugaev}} \bibnamefont{and}
  \bibinfo{author}{\bibfnamefont{P.}~\bibnamefont{Klimai}},
  \bibinfo{journal}{Phys. Rev.} \textbf{\bibinfo{volume}{D90}},
  \bibinfo{pages}{103501} (\bibinfo{year}{2014}), \eprint{1312.7435}.

\bibitem[{\citenamefont{Kim}(1979)}]{Kim:1979if}
\bibinfo{author}{\bibfnamefont{J.~E.} \bibnamefont{Kim}},
  \bibinfo{journal}{Phys. Rev. Lett.} \textbf{\bibinfo{volume}{43}},
  \bibinfo{pages}{103} (\bibinfo{year}{1979}).

\bibitem[{\citenamefont{Shifman et~al.}(1980)\citenamefont{Shifman, Vainshtein,
  and Zakharov}}]{Shifman:1979if}
\bibinfo{author}{\bibfnamefont{M.~A.} \bibnamefont{Shifman}},
  \bibinfo{author}{\bibfnamefont{A.~I.} \bibnamefont{Vainshtein}},
  \bibnamefont{and} \bibinfo{author}{\bibfnamefont{V.~I.}
  \bibnamefont{Zakharov}}, \bibinfo{journal}{Nucl. Phys.}
  \textbf{\bibinfo{volume}{B166}}, \bibinfo{pages}{493} (\bibinfo{year}{1980}).

\bibitem[{\citenamefont{Zhitnitsky}(1980)}]{Zhitnitsky:1980tq}
\bibinfo{author}{\bibfnamefont{A.~R.} \bibnamefont{Zhitnitsky}},
  \bibinfo{journal}{Sov. J. Nucl. Phys.} \textbf{\bibinfo{volume}{31}},
  \bibinfo{pages}{260} (\bibinfo{year}{1980}), \bibinfo{note}{[Yad.
  Fiz.31,497(1980)]}.

\bibitem[{\citenamefont{Dine et~al.}(1981)\citenamefont{Dine, Fischler, and
  Srednicki}}]{Dine:1981rt}
\bibinfo{author}{\bibfnamefont{M.}~\bibnamefont{Dine}},
  \bibinfo{author}{\bibfnamefont{W.}~\bibnamefont{Fischler}}, \bibnamefont{and}
  \bibinfo{author}{\bibfnamefont{M.}~\bibnamefont{Srednicki}},
  \bibinfo{journal}{Phys. Lett.} \textbf{\bibinfo{volume}{104B}},
  \bibinfo{pages}{199} (\bibinfo{year}{1981}).

\bibitem[{\citenamefont{Okada and Raut}(2017)}]{Okada:2016ssd}
\bibinfo{author}{\bibfnamefont{N.}~\bibnamefont{Okada}} \bibnamefont{and}
  \bibinfo{author}{\bibfnamefont{D.}~\bibnamefont{Raut}},
  \bibinfo{journal}{Phys. Rev.} \textbf{\bibinfo{volume}{D95}},
  \bibinfo{pages}{035035} (\bibinfo{year}{2017}), \eprint{1610.09362}.

\bibitem[{\citenamefont{Okada et~al.}(2017)\citenamefont{Okada, Okada, and
  Raut}}]{Okada:2017cvy}
\bibinfo{author}{\bibfnamefont{N.}~\bibnamefont{Okada}},
  \bibinfo{author}{\bibfnamefont{S.}~\bibnamefont{Okada}}, \bibnamefont{and}
  \bibinfo{author}{\bibfnamefont{D.}~\bibnamefont{Raut}},
  \bibinfo{journal}{Phys. Rev.} \textbf{\bibinfo{volume}{D95}},
  \bibinfo{pages}{055030} (\bibinfo{year}{2017}), \eprint{1702.02938}.

\bibitem[{\citenamefont{Choi and Lee}(2016)}]{Choi:2016eif}
\bibinfo{author}{\bibfnamefont{S.-M.} \bibnamefont{Choi}} \bibnamefont{and}
  \bibinfo{author}{\bibfnamefont{H.~M.} \bibnamefont{Lee}},
  \bibinfo{journal}{Eur. Phys. J.} \textbf{\bibinfo{volume}{C76}},
  \bibinfo{pages}{303} (\bibinfo{year}{2016}), \eprint{1601.05979}.

\bibitem[{\citenamefont{Enqvist et~al.}(2010)\citenamefont{Enqvist, Mazumdar,
  and Stephens}}]{Enqvist:2010vd}
\bibinfo{author}{\bibfnamefont{K.}~\bibnamefont{Enqvist}},
  \bibinfo{author}{\bibfnamefont{A.}~\bibnamefont{Mazumdar}}, \bibnamefont{and}
  \bibinfo{author}{\bibfnamefont{P.}~\bibnamefont{Stephens}},
  \bibinfo{journal}{JCAP} \textbf{\bibinfo{volume}{1006}}, \bibinfo{pages}{020}
  (\bibinfo{year}{2010}), \eprint{1004.3724}.

\bibitem[{\citenamefont{Coleman and Weinberg}(1973)}]{Coleman:1973jx}
\bibinfo{author}{\bibfnamefont{S.~R.} \bibnamefont{Coleman}} \bibnamefont{and}
  \bibinfo{author}{\bibfnamefont{E.~J.} \bibnamefont{Weinberg}},
  \bibinfo{journal}{Phys. Rev.} \textbf{\bibinfo{volume}{D7}},
  \bibinfo{pages}{1888} (\bibinfo{year}{1973}).

\bibitem[{\citenamefont{Barenboim et~al.}(2014)\citenamefont{Barenboim, Chun,
  and Lee}}]{Barenboim:2013wra}
\bibinfo{author}{\bibfnamefont{G.}~\bibnamefont{Barenboim}},
  \bibinfo{author}{\bibfnamefont{E.~J.} \bibnamefont{Chun}}, \bibnamefont{and}
  \bibinfo{author}{\bibfnamefont{H.~M.} \bibnamefont{Lee}},
  \bibinfo{journal}{Phys. Lett.} \textbf{\bibinfo{volume}{B730}},
  \bibinfo{pages}{81} (\bibinfo{year}{2014}), \eprint{1309.1695}.

\bibitem[{\citenamefont{Casas et~al.}(1999)\citenamefont{Casas, Di~Clemente,
  and Quiros}}]{Casas:1998cf}
\bibinfo{author}{\bibfnamefont{J.~A.} \bibnamefont{Casas}},
  \bibinfo{author}{\bibfnamefont{V.}~\bibnamefont{Di~Clemente}},
  \bibnamefont{and} \bibinfo{author}{\bibfnamefont{M.}~\bibnamefont{Quiros}},
  \bibinfo{journal}{Nucl. Phys.} \textbf{\bibinfo{volume}{B553}},
  \bibinfo{pages}{511} (\bibinfo{year}{1999}), \eprint{hep-ph/9809275}.

\bibitem[{\citenamefont{Staub}(2014)}]{Staub:2013tta}
\bibinfo{author}{\bibfnamefont{F.}~\bibnamefont{Staub}},
  \bibinfo{journal}{Comput. Phys. Commun.} \textbf{\bibinfo{volume}{185}},
  \bibinfo{pages}{1773} (\bibinfo{year}{2014}), \eprint{1309.7223}.

\bibitem[{\citenamefont{Ballesteros and Casas}(2015)}]{Ballesteros:2014yva}
\bibinfo{author}{\bibfnamefont{G.}~\bibnamefont{Ballesteros}} \bibnamefont{and}
  \bibinfo{author}{\bibfnamefont{J.~A.} \bibnamefont{Casas}},
  \bibinfo{journal}{Phys. Rev.} \textbf{\bibinfo{volume}{D91}},
  \bibinfo{pages}{043502} (\bibinfo{year}{2015}), \eprint{1406.3342}.

\bibitem[{\citenamefont{Raffelt}(1999)}]{Raffelt:1999tx}
\bibinfo{author}{\bibfnamefont{G.~G.} \bibnamefont{Raffelt}},
  \bibinfo{journal}{Ann. Rev. Nucl. Part. Sci.} \textbf{\bibinfo{volume}{49}},
  \bibinfo{pages}{163} (\bibinfo{year}{1999}), \eprint{hep-ph/9903472}.

\bibitem[{\citenamefont{Griest and Kamionkowski}(1990)}]{Griest:1989wd}
\bibinfo{author}{\bibfnamefont{K.}~\bibnamefont{Griest}} \bibnamefont{and}
  \bibinfo{author}{\bibfnamefont{M.}~\bibnamefont{Kamionkowski}},
  \bibinfo{journal}{Phys. Rev. Lett.} \textbf{\bibinfo{volume}{64}},
  \bibinfo{pages}{615} (\bibinfo{year}{1990}).

\bibitem[{\citenamefont{Akhmedov et~al.}(1998)\citenamefont{Akhmedov, Rubakov,
  and Smirnov}}]{Akhmedov:1998qx}
\bibinfo{author}{\bibfnamefont{E.~K.} \bibnamefont{Akhmedov}},
  \bibinfo{author}{\bibfnamefont{V.~A.} \bibnamefont{Rubakov}},
  \bibnamefont{and} \bibinfo{author}{\bibfnamefont{A.~{\relax Yu}.}
  \bibnamefont{Smirnov}}, \bibinfo{journal}{Phys. Rev. Lett.}
  \textbf{\bibinfo{volume}{81}}, \bibinfo{pages}{1359} (\bibinfo{year}{1998}),
  \eprint{hep-ph/9803255}.

\bibitem[{\citenamefont{Asaka and Shaposhnikov}(2005)}]{Asaka:2005pn}
\bibinfo{author}{\bibfnamefont{T.}~\bibnamefont{Asaka}} \bibnamefont{and}
  \bibinfo{author}{\bibfnamefont{M.}~\bibnamefont{Shaposhnikov}},
  \bibinfo{journal}{Phys. Lett.} \textbf{\bibinfo{volume}{B620}},
  \bibinfo{pages}{17} (\bibinfo{year}{2005}), \eprint{hep-ph/0505013}.

\bibitem[{\citenamefont{Shaposhnikov}(2008)}]{Shaposhnikov:2008pf}
\bibinfo{author}{\bibfnamefont{M.}~\bibnamefont{Shaposhnikov}},
  \bibinfo{journal}{JHEP} \textbf{\bibinfo{volume}{08}}, \bibinfo{pages}{008}
  (\bibinfo{year}{2008}), \eprint{0804.4542}.

\bibitem[{\citenamefont{Canetti et~al.}(2012)\citenamefont{Canetti, Drewes, and
  Shaposhnikov}}]{Canetti:2012zc}
\bibinfo{author}{\bibfnamefont{L.}~\bibnamefont{Canetti}},
  \bibinfo{author}{\bibfnamefont{M.}~\bibnamefont{Drewes}}, \bibnamefont{and}
  \bibinfo{author}{\bibfnamefont{M.}~\bibnamefont{Shaposhnikov}},
  \bibinfo{journal}{New J. Phys.} \textbf{\bibinfo{volume}{14}},
  \bibinfo{pages}{095012} (\bibinfo{year}{2012}), \eprint{1204.4186}.

\bibitem[{\citenamefont{Canetti et~al.}(2013)\citenamefont{Canetti, Drewes,
  Frossard, and Shaposhnikov}}]{Canetti:2012kh}
\bibinfo{author}{\bibfnamefont{L.}~\bibnamefont{Canetti}},
  \bibinfo{author}{\bibfnamefont{M.}~\bibnamefont{Drewes}},
  \bibinfo{author}{\bibfnamefont{T.}~\bibnamefont{Frossard}}, \bibnamefont{and}
  \bibinfo{author}{\bibfnamefont{M.}~\bibnamefont{Shaposhnikov}},
  \bibinfo{journal}{Phys. Rev.} \textbf{\bibinfo{volume}{D87}},
  \bibinfo{pages}{093006} (\bibinfo{year}{2013}), \eprint{1208.4607}.

\bibitem[{\citenamefont{Asaka et~al.}(2012)\citenamefont{Asaka, Eijima, and
  Ishida}}]{Asaka:2011wq}
\bibinfo{author}{\bibfnamefont{T.}~\bibnamefont{Asaka}},
  \bibinfo{author}{\bibfnamefont{S.}~\bibnamefont{Eijima}}, \bibnamefont{and}
  \bibinfo{author}{\bibfnamefont{H.}~\bibnamefont{Ishida}},
  \bibinfo{journal}{JCAP} \textbf{\bibinfo{volume}{1202}}, \bibinfo{pages}{021}
  (\bibinfo{year}{2012}), \eprint{1112.5565}.

\bibitem[{\citenamefont{Shuve and Yavin}(2014)}]{Shuve:2014zua}
\bibinfo{author}{\bibfnamefont{B.}~\bibnamefont{Shuve}} \bibnamefont{and}
  \bibinfo{author}{\bibfnamefont{I.}~\bibnamefont{Yavin}},
  \bibinfo{journal}{Phys. Rev.} \textbf{\bibinfo{volume}{D89}},
  \bibinfo{pages}{075014} (\bibinfo{year}{2014}), \eprint{1401.2459}.

\bibitem[{\citenamefont{Abada et~al.}(2015{\natexlab{a}})\citenamefont{Abada,
  Arcadi, Domcke, and Lucente}}]{Abada:2015rta}
\bibinfo{author}{\bibfnamefont{A.}~\bibnamefont{Abada}},
  \bibinfo{author}{\bibfnamefont{G.}~\bibnamefont{Arcadi}},
  \bibinfo{author}{\bibfnamefont{V.}~\bibnamefont{Domcke}}, \bibnamefont{and}
  \bibinfo{author}{\bibfnamefont{M.}~\bibnamefont{Lucente}},
  \bibinfo{journal}{JCAP} \textbf{\bibinfo{volume}{1511}}, \bibinfo{pages}{041}
  (\bibinfo{year}{2015}{\natexlab{a}}), \eprint{1507.06215}.

\bibitem[{\citenamefont{Hernandez et~al.}(2015)\citenamefont{Hernandez, Kekic,
  Lopez-Pavon, Racker, and Rius}}]{Hernandez:2015wna}
\bibinfo{author}{\bibfnamefont{P.}~\bibnamefont{Hernandez}},
  \bibinfo{author}{\bibfnamefont{M.}~\bibnamefont{Kekic}},
  \bibinfo{author}{\bibfnamefont{J.}~\bibnamefont{Lopez-Pavon}},
  \bibinfo{author}{\bibfnamefont{J.}~\bibnamefont{Racker}}, \bibnamefont{and}
  \bibinfo{author}{\bibfnamefont{N.}~\bibnamefont{Rius}},
  \bibinfo{journal}{JHEP} \textbf{\bibinfo{volume}{10}}, \bibinfo{pages}{067}
  (\bibinfo{year}{2015}), \eprint{1508.03676}.

\bibitem[{\citenamefont{Hernandez et~al.}(2016)\citenamefont{Hernandez, Kekic,
  Lopez-Pavon, Racker, and Salvado}}]{Hernandez:2016kel}
\bibinfo{author}{\bibfnamefont{P.}~\bibnamefont{Hernandez}},
  \bibinfo{author}{\bibfnamefont{M.}~\bibnamefont{Kekic}},
  \bibinfo{author}{\bibfnamefont{J.}~\bibnamefont{Lopez-Pavon}},
  \bibinfo{author}{\bibfnamefont{J.}~\bibnamefont{Racker}}, \bibnamefont{and}
  \bibinfo{author}{\bibfnamefont{J.}~\bibnamefont{Salvado}},
  \bibinfo{journal}{JHEP} \textbf{\bibinfo{volume}{08}}, \bibinfo{pages}{157}
  (\bibinfo{year}{2016}), \eprint{1606.06719}.

\bibitem[{\citenamefont{Drewes et~al.}(2016)\citenamefont{Drewes, Garbrecht,
  Gueter, and Klaric}}]{Drewes:2016gmt}
\bibinfo{author}{\bibfnamefont{M.}~\bibnamefont{Drewes}},
  \bibinfo{author}{\bibfnamefont{B.}~\bibnamefont{Garbrecht}},
  \bibinfo{author}{\bibfnamefont{D.}~\bibnamefont{Gueter}}, \bibnamefont{and}
  \bibinfo{author}{\bibfnamefont{J.}~\bibnamefont{Klaric}},
  \bibinfo{journal}{JHEP} \textbf{\bibinfo{volume}{12}}, \bibinfo{pages}{150}
  (\bibinfo{year}{2016}), \eprint{1606.06690}.

\bibitem[{\citenamefont{Drewes et~al.}(2017)\citenamefont{Drewes, Garbrecht,
  Gueter, and Klaric}}]{Drewes:2016jae}
\bibinfo{author}{\bibfnamefont{M.}~\bibnamefont{Drewes}},
  \bibinfo{author}{\bibfnamefont{B.}~\bibnamefont{Garbrecht}},
  \bibinfo{author}{\bibfnamefont{D.}~\bibnamefont{Gueter}}, \bibnamefont{and}
  \bibinfo{author}{\bibfnamefont{J.}~\bibnamefont{Klaric}},
  \bibinfo{journal}{JHEP} \textbf{\bibinfo{volume}{08}}, \bibinfo{pages}{018}
  (\bibinfo{year}{2017}), \eprint{1609.09069}.

\bibitem[{\citenamefont{Hambye and Teresi}(2016)}]{Hambye:2016sby}
\bibinfo{author}{\bibfnamefont{T.}~\bibnamefont{Hambye}} \bibnamefont{and}
  \bibinfo{author}{\bibfnamefont{D.}~\bibnamefont{Teresi}},
  \bibinfo{journal}{Phys. Rev. Lett.} \textbf{\bibinfo{volume}{117}},
  \bibinfo{pages}{091801} (\bibinfo{year}{2016}), \eprint{1606.00017}.

\bibitem[{\citenamefont{Ghiglieri and Laine}(2017)}]{Ghiglieri:2017gjz}
\bibinfo{author}{\bibfnamefont{J.}~\bibnamefont{Ghiglieri}} \bibnamefont{and}
  \bibinfo{author}{\bibfnamefont{M.}~\bibnamefont{Laine}},
  \bibinfo{journal}{JHEP} \textbf{\bibinfo{volume}{05}}, \bibinfo{pages}{132}
  (\bibinfo{year}{2017}), \eprint{1703.06087}.

\bibitem[{\citenamefont{Asaka et~al.}(2017)\citenamefont{Asaka, Eijima, Ishida,
  Minogawa, and Yoshii}}]{Asaka:2017rdj}
\bibinfo{author}{\bibfnamefont{T.}~\bibnamefont{Asaka}},
  \bibinfo{author}{\bibfnamefont{S.}~\bibnamefont{Eijima}},
  \bibinfo{author}{\bibfnamefont{H.}~\bibnamefont{Ishida}},
  \bibinfo{author}{\bibfnamefont{K.}~\bibnamefont{Minogawa}}, \bibnamefont{and}
  \bibinfo{author}{\bibfnamefont{T.}~\bibnamefont{Yoshii}}
  (\bibinfo{year}{2017}), \eprint{1704.02692}.

\bibitem[{\citenamefont{Hambye and Teresi}(2017)}]{Hambye:2017elz}
\bibinfo{author}{\bibfnamefont{T.}~\bibnamefont{Hambye}} \bibnamefont{and}
  \bibinfo{author}{\bibfnamefont{D.}~\bibnamefont{Teresi}},
  \bibinfo{journal}{Phys. Rev.} \textbf{\bibinfo{volume}{D96}},
  \bibinfo{pages}{015031} (\bibinfo{year}{2017}), \eprint{1705.00016}.

\bibitem[{\citenamefont{Abada et~al.}(2017)\citenamefont{Abada, Arcadi, Domcke,
  and Lucente}}]{Abada:2017ieq}
\bibinfo{author}{\bibfnamefont{A.}~\bibnamefont{Abada}},
  \bibinfo{author}{\bibfnamefont{G.}~\bibnamefont{Arcadi}},
  \bibinfo{author}{\bibfnamefont{V.}~\bibnamefont{Domcke}}, \bibnamefont{and}
  \bibinfo{author}{\bibfnamefont{M.}~\bibnamefont{Lucente}},
  \bibinfo{journal}{JCAP} \textbf{\bibinfo{volume}{1712}}, \bibinfo{pages}{024}
  (\bibinfo{year}{2017}), \eprint{1709.00415}.

\bibitem[{\citenamefont{Ghiglieri and Laine}(2018)}]{Ghiglieri:2017csp}
\bibinfo{author}{\bibfnamefont{J.}~\bibnamefont{Ghiglieri}} \bibnamefont{and}
  \bibinfo{author}{\bibfnamefont{M.}~\bibnamefont{Laine}},
  \bibinfo{journal}{JHEP} \textbf{\bibinfo{volume}{02}}, \bibinfo{pages}{078}
  (\bibinfo{year}{2018}), \eprint{1711.08469}.

\bibitem[{\citenamefont{Ferrari et~al.}(2000)\citenamefont{Ferrari, Collot,
  Andrieux, Belhorma, de~Saintignon, Hostachy, Martin, and
  Wielers}}]{Ferrari:2000sp}
\bibinfo{author}{\bibfnamefont{A.}~\bibnamefont{Ferrari}},
  \bibinfo{author}{\bibfnamefont{J.}~\bibnamefont{Collot}},
  \bibinfo{author}{\bibfnamefont{M.-L.} \bibnamefont{Andrieux}},
  \bibinfo{author}{\bibfnamefont{B.}~\bibnamefont{Belhorma}},
  \bibinfo{author}{\bibfnamefont{P.}~\bibnamefont{de~Saintignon}},
  \bibinfo{author}{\bibfnamefont{J.-Y.} \bibnamefont{Hostachy}},
  \bibinfo{author}{\bibfnamefont{P.}~\bibnamefont{Martin}}, \bibnamefont{and}
  \bibinfo{author}{\bibfnamefont{M.}~\bibnamefont{Wielers}},
  \bibinfo{journal}{Phys. Rev.} \textbf{\bibinfo{volume}{D62}},
  \bibinfo{pages}{013001} (\bibinfo{year}{2000}).

\bibitem[{\citenamefont{Graesser}(2007)}]{Graesser:2007pc}
\bibinfo{author}{\bibfnamefont{M.~L.} \bibnamefont{Graesser}}
  (\bibinfo{year}{2007}), \eprint{0705.2190}.

\bibitem[{\citenamefont{del Aguila and
  Aguilar-Saavedra}(2009)}]{delAguila:2008cj}
\bibinfo{author}{\bibfnamefont{F.}~\bibnamefont{del Aguila}} \bibnamefont{and}
  \bibinfo{author}{\bibfnamefont{J.~A.} \bibnamefont{Aguilar-Saavedra}},
  \bibinfo{journal}{Nucl. Phys.} \textbf{\bibinfo{volume}{B813}},
  \bibinfo{pages}{22} (\bibinfo{year}{2009}), \eprint{0808.2468}.

\bibitem[{\citenamefont{Bhupal~Dev et~al.}(2012)\citenamefont{Bhupal~Dev,
  Franceschini, and Mohapatra}}]{BhupalDev:2012zg}
\bibinfo{author}{\bibfnamefont{P.~S.} \bibnamefont{Bhupal~Dev}},
  \bibinfo{author}{\bibfnamefont{R.}~\bibnamefont{Franceschini}},
  \bibnamefont{and} \bibinfo{author}{\bibfnamefont{R.~N.}
  \bibnamefont{Mohapatra}}, \bibinfo{journal}{Phys. Rev.}
  \textbf{\bibinfo{volume}{D86}}, \bibinfo{pages}{093010}
  (\bibinfo{year}{2012}), \eprint{1207.2756}.

\bibitem[{\citenamefont{Helo et~al.}(2014)\citenamefont{Helo, Hirsch, and
  Kovalenko}}]{Helo:2013esa}
\bibinfo{author}{\bibfnamefont{J.~C.} \bibnamefont{Helo}},
  \bibinfo{author}{\bibfnamefont{M.}~\bibnamefont{Hirsch}}, \bibnamefont{and}
  \bibinfo{author}{\bibfnamefont{S.}~\bibnamefont{Kovalenko}},
  \bibinfo{journal}{Phys. Rev.} \textbf{\bibinfo{volume}{D89}},
  \bibinfo{pages}{073005} (\bibinfo{year}{2014}), \bibinfo{note}{[Erratum:
  Phys. Rev.D93,no.9,099902(2016)]}, \eprint{1312.2900}.

\bibitem[{\citenamefont{Blondel et~al.}(2016)\citenamefont{Blondel, Graverini,
  Serra, and Shaposhnikov}}]{Blondel:2014bra}
\bibinfo{author}{\bibfnamefont{A.}~\bibnamefont{Blondel}},
  \bibinfo{author}{\bibfnamefont{E.}~\bibnamefont{Graverini}},
  \bibinfo{author}{\bibfnamefont{N.}~\bibnamefont{Serra}}, \bibnamefont{and}
  \bibinfo{author}{\bibfnamefont{M.}~\bibnamefont{Shaposhnikov}}
  (\bibinfo{collaboration}{FCC-ee study Team}), \bibinfo{journal}{Nucl. Part.
  Phys. Proc.} \textbf{\bibinfo{volume}{273-275}}, \bibinfo{pages}{1883}
  (\bibinfo{year}{2016}), \eprint{1411.5230}.

\bibitem[{\citenamefont{Abada et~al.}(2015{\natexlab{b}})\citenamefont{Abada,
  De~Romeri, Monteil, Orloff, and Teixeira}}]{Abada:2014cca}
\bibinfo{author}{\bibfnamefont{A.}~\bibnamefont{Abada}},
  \bibinfo{author}{\bibfnamefont{V.}~\bibnamefont{De~Romeri}},
  \bibinfo{author}{\bibfnamefont{S.}~\bibnamefont{Monteil}},
  \bibinfo{author}{\bibfnamefont{J.}~\bibnamefont{Orloff}}, \bibnamefont{and}
  \bibinfo{author}{\bibfnamefont{A.~M.} \bibnamefont{Teixeira}},
  \bibinfo{journal}{JHEP} \textbf{\bibinfo{volume}{04}}, \bibinfo{pages}{051}
  (\bibinfo{year}{2015}{\natexlab{b}}), \eprint{1412.6322}.

\bibitem[{\citenamefont{Cui and Shuve}(2015)}]{Cui:2014twa}
\bibinfo{author}{\bibfnamefont{Y.}~\bibnamefont{Cui}} \bibnamefont{and}
  \bibinfo{author}{\bibfnamefont{B.}~\bibnamefont{Shuve}},
  \bibinfo{journal}{JHEP} \textbf{\bibinfo{volume}{02}}, \bibinfo{pages}{049}
  (\bibinfo{year}{2015}), \eprint{1409.6729}.

\bibitem[{\citenamefont{Antusch and Fischer}(2015)}]{Antusch:2015mia}
\bibinfo{author}{\bibfnamefont{S.}~\bibnamefont{Antusch}} \bibnamefont{and}
  \bibinfo{author}{\bibfnamefont{O.}~\bibnamefont{Fischer}},
  \bibinfo{journal}{JHEP} \textbf{\bibinfo{volume}{05}}, \bibinfo{pages}{053}
  (\bibinfo{year}{2015}), \eprint{1502.05915}.

\bibitem[{\citenamefont{Gago et~al.}(2015)\citenamefont{Gago, Hernandez,
  Jones-Perez, Losada, and Moreno~Briceño}}]{Gago:2015vma}
\bibinfo{author}{\bibfnamefont{A.~M.} \bibnamefont{Gago}},
  \bibinfo{author}{\bibfnamefont{P.}~\bibnamefont{Hernandez}},
  \bibinfo{author}{\bibfnamefont{J.}~\bibnamefont{Jones-Perez}},
  \bibinfo{author}{\bibfnamefont{M.}~\bibnamefont{Losada}}, \bibnamefont{and}
  \bibinfo{author}{\bibfnamefont{A.}~\bibnamefont{Moreno~Briceño}},
  \bibinfo{journal}{Eur. Phys. J.} \textbf{\bibinfo{volume}{C75}},
  \bibinfo{pages}{470} (\bibinfo{year}{2015}), \eprint{1505.05880}.

\bibitem[{\citenamefont{Antusch et~al.}(2016)\citenamefont{Antusch, Cazzato,
  and Fischer}}]{Antusch:2016vyf}
\bibinfo{author}{\bibfnamefont{S.}~\bibnamefont{Antusch}},
  \bibinfo{author}{\bibfnamefont{E.}~\bibnamefont{Cazzato}}, \bibnamefont{and}
  \bibinfo{author}{\bibfnamefont{O.}~\bibnamefont{Fischer}},
  \bibinfo{journal}{JHEP} \textbf{\bibinfo{volume}{12}}, \bibinfo{pages}{007}
  (\bibinfo{year}{2016}), \eprint{1604.02420}.

\bibitem[{\citenamefont{Caputo et~al.}(2017{\natexlab{a}})\citenamefont{Caputo,
  Hernandez, Kekic, Lopez-Pavon, and Salvado}}]{Caputo:2016ojx}
\bibinfo{author}{\bibfnamefont{A.}~\bibnamefont{Caputo}},
  \bibinfo{author}{\bibfnamefont{P.}~\bibnamefont{Hernandez}},
  \bibinfo{author}{\bibfnamefont{M.}~\bibnamefont{Kekic}},
  \bibinfo{author}{\bibfnamefont{J.}~\bibnamefont{Lopez-Pavon}},
  \bibnamefont{and} \bibinfo{author}{\bibfnamefont{J.}~\bibnamefont{Salvado}},
  \bibinfo{journal}{Eur. Phys. J.} \textbf{\bibinfo{volume}{C77}},
  \bibinfo{pages}{258} (\bibinfo{year}{2017}{\natexlab{a}}),
  \eprint{1611.05000}.

\bibitem[{\citenamefont{Caputo et~al.}(2017{\natexlab{b}})\citenamefont{Caputo,
  Hernandez, Lopez-Pavon, and Salvado}}]{Caputo:2017pit}
\bibinfo{author}{\bibfnamefont{A.}~\bibnamefont{Caputo}},
  \bibinfo{author}{\bibfnamefont{P.}~\bibnamefont{Hernandez}},
  \bibinfo{author}{\bibfnamefont{J.}~\bibnamefont{Lopez-Pavon}},
  \bibnamefont{and} \bibinfo{author}{\bibfnamefont{J.}~\bibnamefont{Salvado}},
  \bibinfo{journal}{JHEP} \textbf{\bibinfo{volume}{06}}, \bibinfo{pages}{112}
  (\bibinfo{year}{2017}{\natexlab{b}}), \eprint{1704.08721}.

\bibitem[{\citenamefont{Langacker et~al.}(1986)\citenamefont{Langacker, Peccei,
  and Yanagida}}]{Langacker:1986rj}
\bibinfo{author}{\bibfnamefont{P.}~\bibnamefont{Langacker}},
  \bibinfo{author}{\bibfnamefont{R.~D.} \bibnamefont{Peccei}},
  \bibnamefont{and} \bibinfo{author}{\bibfnamefont{T.}~\bibnamefont{Yanagida}},
  \bibinfo{journal}{Mod. Phys. Lett.} \textbf{\bibinfo{volume}{A1}},
  \bibinfo{pages}{541} (\bibinfo{year}{1986}).

\bibitem[{\citenamefont{Caputo et~al.}(2018)\citenamefont{Caputo, Hernandez,
  and Rius}}]{Caputo:2018zky}
\bibinfo{author}{\bibfnamefont{A.}~\bibnamefont{Caputo}},
  \bibinfo{author}{\bibfnamefont{P.}~\bibnamefont{Hernandez}},
  \bibnamefont{and} \bibinfo{author}{\bibfnamefont{N.}~\bibnamefont{Rius}}
  (\bibinfo{year}{2018}), \eprint{1807.03309}.

\bibitem[{\citenamefont{Clarke and Volkas}(2016)}]{Clarke:2015bea}
\bibinfo{author}{\bibfnamefont{J.~D.} \bibnamefont{Clarke}} \bibnamefont{and}
  \bibinfo{author}{\bibfnamefont{R.~R.} \bibnamefont{Volkas}},
  \bibinfo{journal}{Phys. Rev.} \textbf{\bibinfo{volume}{D93}},
  \bibinfo{pages}{035001} (\bibinfo{year}{2016}), \bibinfo{note}{[Phys.
  Rev.D93,035001(2016)]}, \eprint{1509.07243}.

\end{thebibliography}
\bibliographystyle{apsrev}

\end{document}